\documentclass[aps,pra,twocolumn,amsmath,amssymb,nofootinbib]{revtex4-1}

\usepackage{graphicx}
\usepackage{dcolumn}
\usepackage{braket}
\usepackage{setspace}
\usepackage{bm}
\usepackage{amsmath}
\usepackage{csquotes}
\usepackage{mathrsfs}
\usepackage{titlesec}
\usepackage{subfigure}
\usepackage[svgnames]{xcolor}

\usepackage[colorlinks,citecolor=red,urlcolor=red,bookmarks=false,  hypertexnames=true]{hyperref}

\begin{document}
\title{Spatial Mode Correction of Single Photons using Machine Learning}

\author{Narayan Bhusal$^{1}$}
\email{nbhusa1@lsu.edu}
\author{Sanjaya Lohani$^{2}$}
\author{Chenglong You$^{1}$}
\email{cyou2@lsu.edu}
\author{Mingyuan Hong$^{1}$}
\author{Joshua Fabre$^{1}$}
\author{Pengcheng Zhao$^{3}$}
\author{Erin M. Knutson$^{2}$}
\author{Ryan T. Glasser$^{2}$}
\author{Omar S. Maga\~na-Loaiza$^{1}$}

\affiliation{$^{1}$Department of Physics \& Astronomy, Louisiana State University, Baton Rouge, LA 70803, USA}
\affiliation{$^{2}$Tulane University, New Orleans, LA 70118, USA}
\affiliation{$^{3}$School of Information Science and Technology,\\ Qingdao University of Science and Technology, China}

\date{\today}
\begin{abstract}
Spatial modes of light constitute valuable resources for a variety of quantum technologies ranging from quantum communication and quantum imaging to remote sensing. Nevertheless, their vulnerabilities to phase distortions, induced by random media, impose significant limitations on the realistic implementation of numerous quantum-photonic technologies. Unfortunately, this problem is exacerbated at the single-photon level. Over the last two decades, this challenging problem has been tackled through conventional schemes that utilize optical nonlinearities, quantum correlations, and adaptive optics. In this article, we exploit the self-learning and self-evolving features of artificial neural networks to correct the complex spatial profile of distorted Laguerre-Gaussian modes at the single-photon level. Furthermore, 
we demonstrate the possibility of 
boosting the performance of an optical communication protocol through the spatial mode correction of single photons using machine learning. Our results have important implications for real-time turbulence correction of structured photons and single-photon images.
\end{abstract}

\maketitle
\section{Introduction}
\label{sec:1}
\ \ \ Spatially structured beams of light have been extensively used over the last two decades for multiple applications ranging from 3D surface imaging to quantum cryptography \cite{dunlop:2016,bell:1999,geng:2010}. In this regard, Laguerre-Gaussian (LG) modes represent an important family of spatial modes possessing orbital angular momentum (OAM) \cite{andrew:2012}. The OAM of photons is due to a helical phase front given by an azimuthal phase dependence of the form $e^{i\ell\phi}$, where $\ell$ represents the OAM number and $\phi$ represents the azimuthal angle. These beams have enabled the encoding of many bits of information in a single photon, a possibility that has enabled new communication and encryption protocols \cite{omar:2019,willner:2015,yao:2011,wang:2012,milione:2015}. In the past, these optical modes have been exploited to demonstrate high-speed communication in fiber, free-space, and underwater \cite{krenn2015twisted, baghdady:2016}. Furthermore, structured light beams have enabled increased levels of security against eavesdroppers, a crucial feature for secure communication applications \cite{bouren:2002, mirho:2015, grobla:2006, willner:2015, malik:2012}. Last but not least, structured spatial profiles of single photons have been proven to be extremely useful for remote sensing technologies and correlated imaging \cite{omar:2016, yang2017,cvijetic:2015, pattar:2013, omarm:2019, milione:2017, lavery:2013,chen:2014,jack:2009, malik:2014}.

Unfortunately, the spatial profile of photons can be easily distorted in realistic environments \cite{brandon:2012}. Indeed, random phase distortions and scattering effects can destroy information encoded in structured beams of light \cite{roden:2014, paterson:2005, tyler:2009}. Consequently, these spatial distortions severely degrade the performance of protocols for communication, cryptography, and remote sensing \cite{yao:2011, malik:2012}. These problems are exacerbated at the single-photon level, imposing important limitations on the realistic implementation of quantum photonic technologies. Hitherto, these limitations have been alleviated through conventional schemes that use adaptive optics, quantum correlations, and nonlinear optics \cite{hugo:2018,thomas:2018, nick:2019, roden:2014}. However, an efficient and fast protocol to overcome undesirable turbulence effects, at the single-photon level, has not yet been experimentally demonstrated.

Recently, artificial intelligence has gained popularity in optics due to its unique potential for handling complex classification and optimization tasks \cite{lin:2018, dunjko:2018, lecun:2015, biamonte:2017, sush:2020,cai2015entanglement,taira2020}. Indeed, machine learning has been used to engineer  quantum states of light \cite{Krenn:2016, cui:2019}, and to identify their properties in different degrees of freedom \cite{gao:2018, cyou:2020}. Moreover, convolutional neural networks (CNNs) have been demonstrated to be efficient in learning and characterizing the topographical features of images \cite{gu:2018}. The self-evolving and self-learning features of artificial neural networks have been exploited in quantum optical systems for preparation, classification, and characterization. Remarkably, for these particular tasks, machine learning techniques have outperformed conventional approaches \cite{turpin2018, sanj:2018, sanjaya:2018}.

Here, we demonstrate a smart communication protocol that exploits the self-learning features of convolutional neural networks to correct the spatial profile of single photons. The robustness and efficiency of our scheme is tested in a communication protocol that utilizes LG modes. Our results dramatically outperform previous protocols that rely on conventional adaptive optics \cite{roden:2014, paterson:2005, tyler:2009, hugo:2018}. Furthermore, we demonstrate near-unity corrected fidelity in time periods that are comparable to the fluctuation of atmospheric turbulence. Our results have significant implications for various  technologies that exploit single photons with complex spatial profiles \cite{yao:2011,omar:2019,willner:2015}. In addition, our work shows an enormous potential to enable the possibility of overcoming phase distortions induced by thick atmospheric turbulence in real-time.

\begin{figure*}[!ht]
    \centering
    \includegraphics[width=1.0\textwidth]{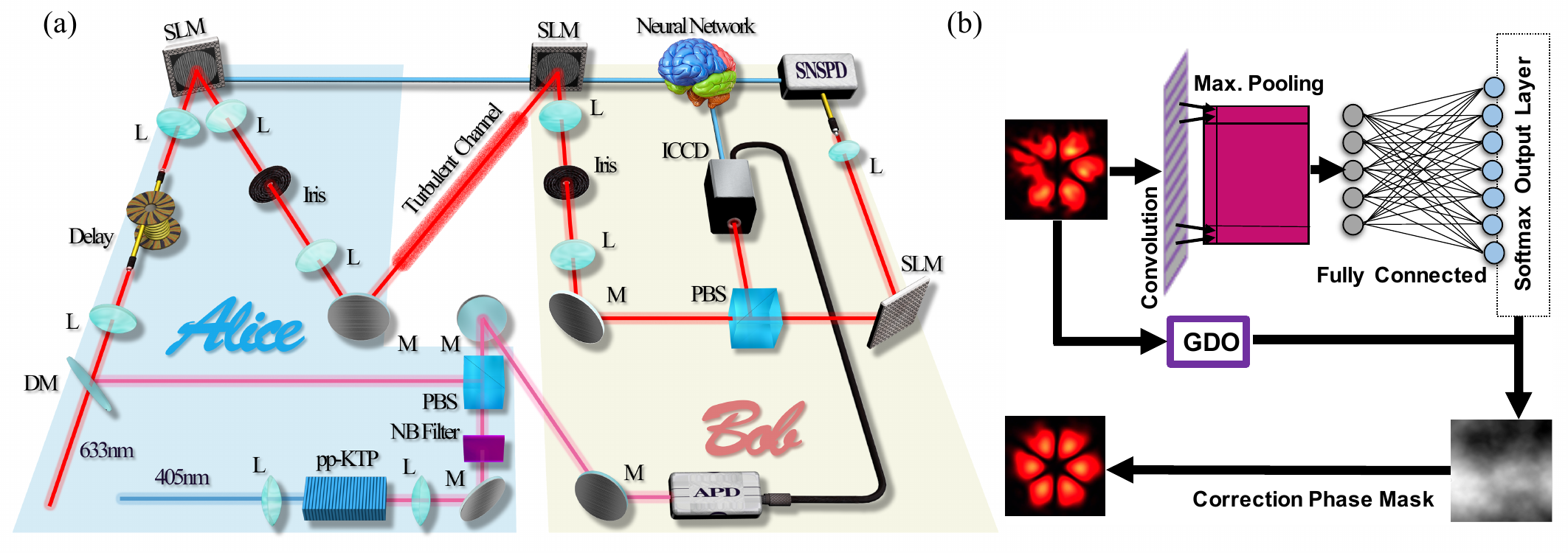}
    \caption{The schematic diagram of the setup used to demonstrate turbulence correction is shown in (a). The experiment is performed using a He-Ne laser and heralded single photons produced through a process of spontaneous parametric down-conversion (SPDC). The sources are switched using a dichroic mirror (DM). The spatial profile of photons is shaped by Alice using a spatial light modulator (SLM). The prepared photons are sent to Bob through a turbulent channel. Bob then performs correction and quantum state tomography on the structured photons. In order to do this, Bob collects multiple classical and single-photon images to train an artificial neural network. The high-light level images are obtained with a CCD camera, whereas single-photon images are formed on a gated ICCD camera. (b) The neural network for turbulence correction comprises a 5-layer convolutional neural network (CNN), and a feedback loop with a gradient descent optimizer (GDO).}
    \label{fig:1}
\end{figure*}

\section{Experimental and Computational Methods}
\label{sec:2}
\ \ \ The schematic diagram of our communication protocol and the computational model of our artificial neural network are depicted in Fig. \ref{fig:1}. Here, Alice prepares spatial modes that are transmitted to Bob through a turbulent communication channel. The atmospheric turbulence in the communication channel induces aberrations in the optical beams that degrade the quality of the information encoded in their phase. This undesirable effect compromises Bob's ability to correctly decode and make measurements on the spatial modes. Bob overcomes this problem by training an artificial neural network with multiple turbulence-distorted beams that allow him to correct the spatial profile of photons.

In our experiment, we use a spatial light modulator (SLM) and computer-generated holograms to produce LG modes. This technique allows us to generate any arbitrary spatial mode in the first-diffraction order of the SLM. The generated modes are filtered and collimated using a 4f-optical system and then projected onto a second SLM. We use this second spatial modulator to display random phase screens that simulate turbulence \cite{roden:2014}. The beam reflected by the second SLM is then split into two beams using a polarizing beam splitter (PBS). The spatial profiles of the beams reflected by the PBS are recorded by a CCD camera. This information is then used to establish a feedback control loop that relies on convolutional neural networks. This process enables the characterization of turbulence in the communication channel. Furthermore, the beam transmitted by the PBS is characterized through quantum state tomography.

Due to the relevance of single-photon imaging for multiple applications \cite{omar:2019}, we also perform a proof-of-principle experiment using heralded single photons produced by a process of spontaneous parametric down-conversion (SPDC). This configuration allows us to demonstrate the potential of our turbulence correction protocol (TCP) at the single-photon level. For this purpose, we utilize a dichroic mirror (DM) to ease the transition from one source to another as shown in Fig. \ref{fig:1}(a). We produce SPDC photons by pumping a type-II potassium titanyl phosphate (ppKTP) crystal with a continuous wave (CW) diode laser at 405 nm. A PBS is used to separate the correlated photon pairs at 810 nm. We utilize temporal correlations to acquire gated images at the single-photon level using an intensified charged coupled device (ICCD) camera. This is performed by adding a delay line to our experiment. Gating the ICCD camera is crucial for the formation of single-photon images \cite{fickler2013real}.

\begin{figure*}[!t]
  \centering
 \includegraphics[width=1.0\textwidth]{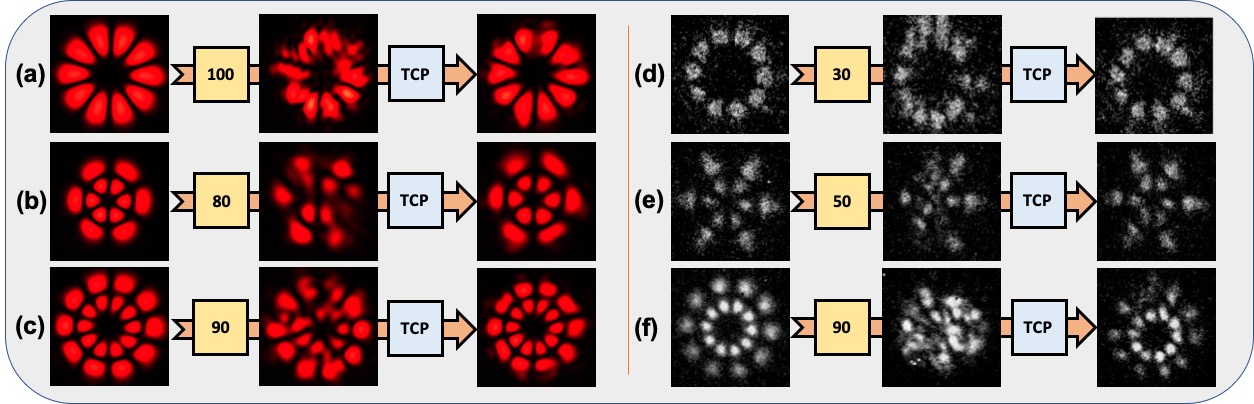}
\caption{Spatial profiles of LG modes at high- and single-photon levels for different turbulence conditions. The first column in each of the panels shows the states prepared by Alice without distortions.
The second columns display the distorted beams measured by Bob. The strength of turbulence is characterized by $C_n^2$ ($\times 10^{-13} \text{mm}^{-2/3})$, these numbers are reported in the yellow rectangle. The corrected spatial profiles are shown in the third column. (a), (b), and (c) show high-light-level demonstrations of our protocol for multiple LG superpositions,  $\ket{\psi}=\frac{1}{\sqrt{2}}\left( \ket{\text{LG}_{+5,0}}+\ket{\text{LG}_{-5,0}}\right)$, $\ket{\psi}=\frac{1}{\sqrt{2}}\left( \ket{\text{LG}_{+3,1}}+\ket{\text{LG}_{-3,1}}\right)$, and $\ket{\psi}=\frac{1}{\sqrt{2}}\left( \ket{\text{LG}_{+5,1}}+\ket{\text{LG}_{-5,1}}\right)$ respectively. The corresponding single-photon demonstrations of (a), (b) and (c) are shown in (d), (e), and (f), respectively.}
\label{fig:2}
\end{figure*}

Fig. \ref{fig:1}(b) illustrates our machine learning algorithm for the correction of structured photons. This is based on a convolutional neural network followed by a gradient descent optimizer \cite{sanjaya:2018, nielsen:2015}. The optimizer consists of a 5-layer CNN and a gradient descent optimization (GDO) algorithm. The CNN takes turbulent images of LG beams, and convolves them with a $5\times5$ filter. The step is immediately followed by a $2\times2$ max-pooling layer before feeding them into 100 fully connected neurons. Finally, the network contains a softmax output layer. We utilize hundreds of realizations of distorted images for multiple turbulence strengths to train the neural networks. The function of the trained CNN is to predict the strength of turbulence in terms of standard refractive index ($C_n^2$) values. The function of the GDO loop is to optimize the correction phase masks over many realizations of random matrices that simulate turbulence. The phase masks are then encoded in the second SLM to obtain the corrected spatial modes at the image plane of the SLM.

We prepare symmetric superpositions of LG modes to demonstrate smart optical communication. This family of modes are solutions to the Helmholtz equation in cylindrical coordinates \cite{omar:2019}. Moreover, these modes form a complete orthonormal basis set with respect to the azimuthal ($\ell$) and the radial ($p$) degrees of freedom \cite{siegman:1986}. In our experiment, we distort the communication modes by using atmospheric turbulence simulated in a SLM \cite{brandon:2016}. We use the Kolmogorov model of turbulence to simulate the turbulent communication channel \cite{roden:2014,sanjaya:2018,Bos:2015}. Turbulence induces a random modulation of the index of refraction that results from inhomogeneities of temperature and pressure of media. This, in turn, leads to distortions of the phase front of the spatial profile of optical modes. The degree of distortion is quantified through the Fried's parameter $r_0$, which is defined in terms of the standard refractive index $C_n^2$,
\begin{equation}
\label{eq:1}
    \Phi(p,q)=\mathbb{R}\left\{\mathcal{F}^{-1}\left(\mathbb{M}_{NN}\sqrt{\phi_{NN}(k)}\right )\right\},
\end{equation}
with $\phi_{NN}(k)=0.023 r_0^{-5/3} \left(k^2+k_0^2\right)^{-11/6} e^{-k^2/k_m^2}$ and the Fried's parameter $r_0=\left(0.423k^2C_n^2d\right)^{-3/5}$. Here, $k$, $d$, and $\mathbb{M}_{NN}$ represent the wave number ($2\pi/\lambda$), the propagation distance, and the encoded random matrix, respectively. Even though the strength of phase distortion can be varied using $d$ and $C_n^2$, we choose to vary its strength using $C_n^2$. Furthermore, we perform the phase mask optimization iteratively using the GDO algorithm
\begin{equation}
\begin{aligned}
\label{eq:2}
\Phi^{j}(p, q)=& \angle\left[\mathcal{F}^{-1}\left\{\frac{1}{H} \times \mathcal{F}\left[\mathcal{F}^{-1}\left(\mathcal{F}\left(G\left(p, q, w_{0}\right)\times\right.\right.\right.\right.\right.\\
&\left.\left.\left.\left.\left.\exp\left(i \Theta^{(\ell,-\ell)}\right)\right) \times H\right) \exp \left(-i \Phi_{\mathrm{est}}^{j}(p, q)\right)\right]\right\}\right].
\end{aligned}
\end{equation}

\noindent The mean squared error (MSE) between the predicted intensity and the corresponding simulated target intensity is used as the cost function. Here, $\Phi^{j}(p,q)$ is the phase mask at the $j^{\text{th}}$ iteration, $G(p,q,w_0)$ represents the Gaussian beam with the waist $w_0$, and $H$ represents the transfer function describing the SLM and the propagation of the beam. The phase mask used to generate the original LG superposition mode is represented by $\Theta^{(\ell,-\ell)}$ in Eq. (\ref{eq:2}).

\section{Results and Discussion}
\label{sec:3}
\ \ \ 

\begin{figure*}[!t]
  \centering
 \includegraphics[width=1.0\textwidth]{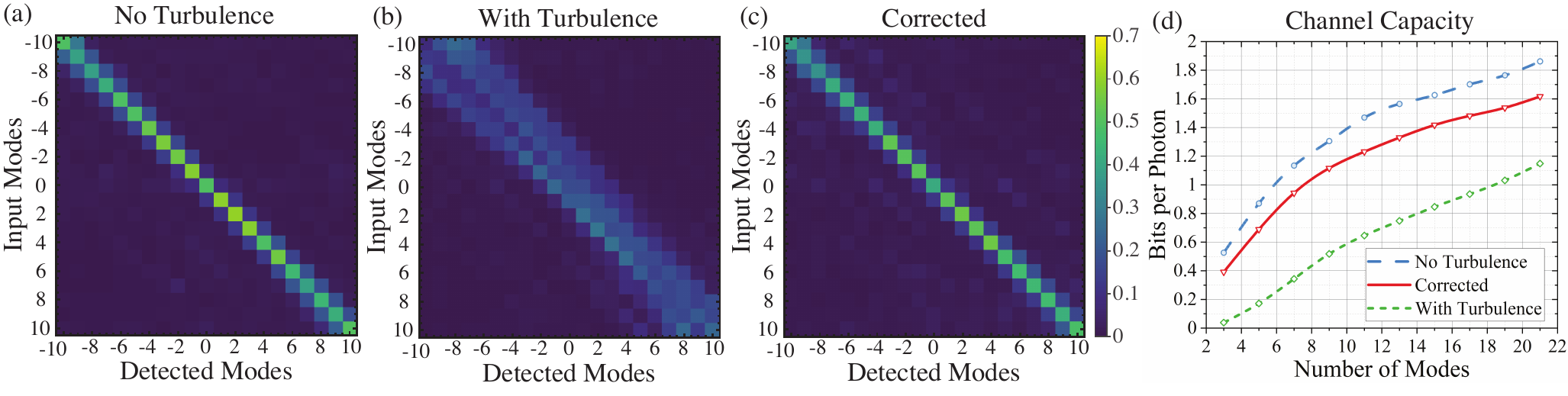}
\caption{The cross-correlation matrices represent conditional probabilities between sent and detected modes in the OAM basis. 
(a) shows the cross-correlation matrix obtained for our communication protocol in the absence of turbulence. In (b), we plot the cross-correlation matrix for our communication protocol in the presence of turbulence characterized by $C_n^2$ = 90$\times 10^{-13} \text{mm}^{-2/3}$. In this case, it is almost impossible to correctly identify the spatial modes. (c) shows the cross-correlation matrix after applying our turbulence correction protocol. Our turbulence correction protocol significantly improves the performance of the communication system. We display channel capacity in terms of bits per photon in (d).}
\label{fig:3}
\end{figure*}

In Fig. \ref{fig:2}(a)-(c), we present experimental results obtained with a He-Ne laser. The first column in each of the panels shows the spatial profile of the undistorted modes prepared by Alice. The spatial profiles of the modes are distorted due to atmospheric turbulence in the communication channel. The aberrated modes are shown in the second column of Fig. \ref{fig:2}. In our experiment, we collected hundreds of realizations of the aberrated beams to train our artificial neural networks. The strength of turbulence predicted by CNN was utilized to perform the phase mask optimization by means of a feedback GDO loop. Thus, the CNN in combination with the GDO loop generates the correction phase masks which are then encoded in the second SLM to alleviate turbulence effects. We indicate this process with the blue box labeled as \enquote{TCP} in Fig. \ref{fig:2}. The CNN was trained in a high-performance computing cluster. We then utilize the pre-trained CNNs, which estimate the turbulence strength and initial phase distribution on the order of milliseconds, together with the GDO on a computer with an Intel(R) Core(TM) i7-8750H CPU $@$ 2.20GHz and 16 GB of RAM to generate optimized turbulence correction phase masks. The corrected intensity profiles measured by Bob are depicted in the last column of each panel. In Fig. \ref{fig:2}(a), we show the spatial profile of a structured beam corrected by our protocol for the superposition of LG modes $\ket{\psi}=\frac{1}{\sqrt{2}}\left( \ket{\text{LG}_{+\ell,0}}+ \ket{\text{LG}_{-\ell,0}}\right)$ with $\ell=5$. In Fig. \ref{fig:2}(b) and \ref{fig:2}(c), we show experimental results for complex LG modes, with radial structure, described by $\ket{\psi}=\frac{1}{\sqrt{2}}\left( \ket{\text{LG}_{+\ell,1}}+ \ket{\text{LG}_{-\ell,1}}\right)$ for $\ell=3$ and $\ell=5$ respectively.
 
We also demonstrate the robustness of our technique to correct the spatial profile of heralded single photons. In Fig. \ref{fig:2}(d), \ref{fig:2}(e), and \ref{fig:2}(f), we display turbulence correction of single photons prepared in LG superpositions with different azimuthal and radial quantum numbers, expressed as $\ket{\psi}=\frac{1}{\sqrt{2}}\left( \ket{\text{LG}_{+5,0}}+\ket{\text{LG}_{-5,0}}\right)$, $\ket{\psi}=\frac{1}{\sqrt{2}}\left( \ket{\text{LG}_{+3,1}}+\ket{\text{LG}_{-3,1}}\right)$, and $\ket{\psi}=\frac{1}{\sqrt{2}}\left( \ket{\text{LG}_{+5,1}}+\ket{\text{LG}_{-5,1}}\right)$ respectively. These images were acquired using an ICCD camera. Each of the background-subtracted images are formed by accumulating photons over a time period of 20 minutes. These images demonstrate an excellent mitigation of the turbulence at the single-photon level.

We quantify the performance of our correction protocol through the channel capacity of our optical communication system. Fig. \ref{fig:3}(a) shows the cross-correlation matrix for different transmitted modes in the absence of turbulence. In order to generate this matrix, Bob performs a series of projective measurements on the modes sent by Alice. The cross-correlation matrix represents the conditional probabilities between the modes sent and detected in the communication protocol. A small spread around the diagonal elements even in the absence of turbulence is caused due to diffraction, the finite size of the optical fibers, and experimental misalignment. The cross-correlation matrix obtained in the presence of atmospheric turbulence is shown in Fig. \ref{fig:3}(b). In this case, the spatial distortion induces modal cross-talk that degrades the performance of the communication protocol. These undesirable effects increase with the strength of turbulence in the communication channel. Indeed, this represents an important limitation of free-space communication with spatial modes of light \cite{omar:2019}. In Fig. \ref{fig:3}(c), we show our experimental results for the cross-correlation matrix after applying our turbulence correction protocol. In this case, the cross-correlation matrix is nearly diagonal, showing a dramatic improvement in the performance of our communication protocol. Furthermore, we calculate the normalized mutual information to quantify the channel capacity in terms of bits per photon \cite{roden:2014} as shown in Fig. \ref{fig:3}(d). We used the conditional probabilities of the cross-correlation matrices to calculate mutual information for Hilbert space dimension $N$ according to the following equation $\text{MI}=\frac{1}{N} \sum_{d,s} P(d \mid s) \log _{2}\left(\frac{P(d \mid s) N}{\sum_{s} P(d \mid s)}\right)$, where the subscripts $d$ and $s$ represent the detected and sent modes respectively. The channel capacity plot demonstrates the potential of our technique to correct spatial modes of light. 

\begin{figure}[!t]
   \centering
  \includegraphics[width=0.4\textwidth]{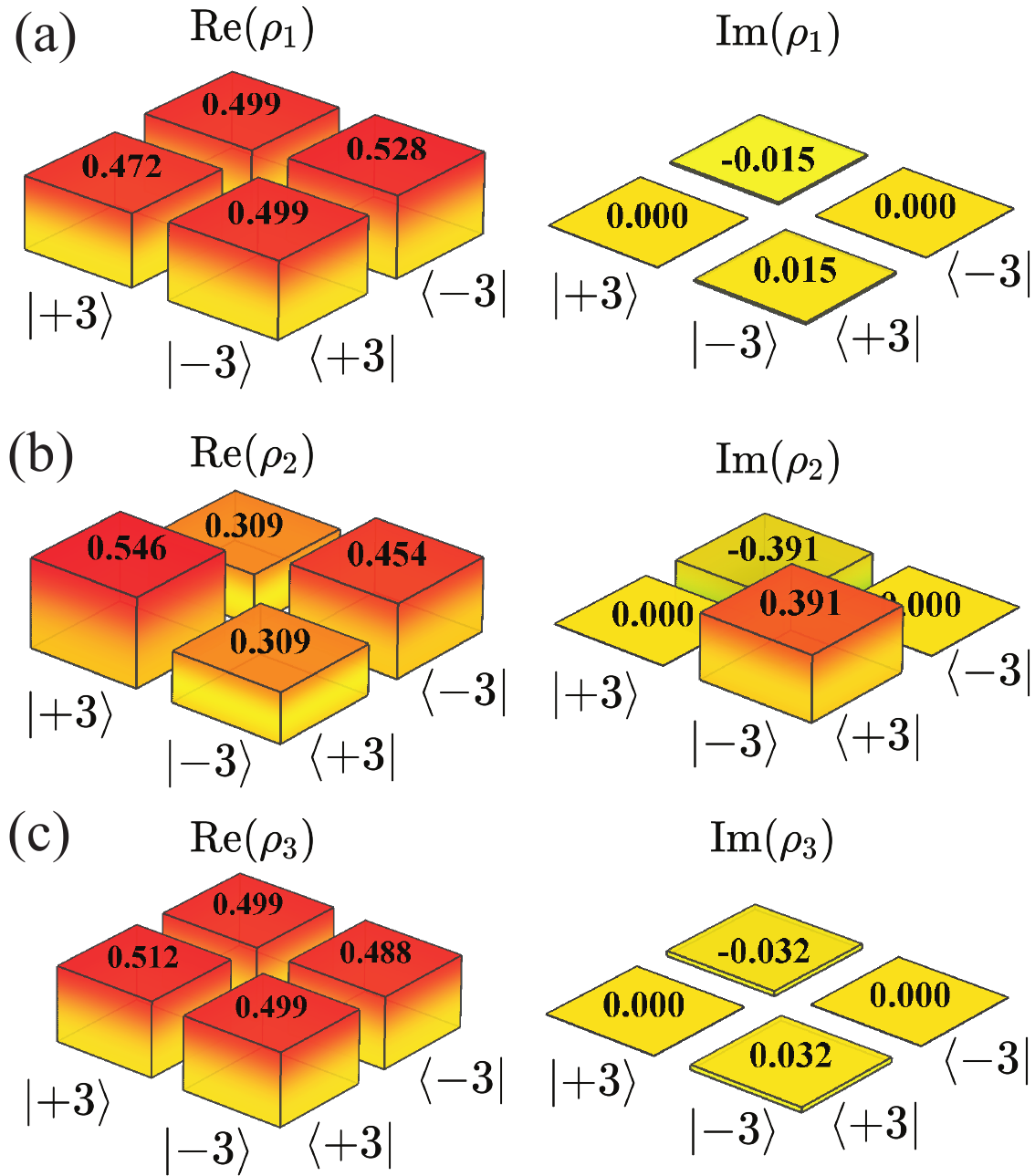}
 \caption{Real and imaginary parts of the density matrices for the qubits encoded in the OAM basis. In this case, we prepared $\ket{\psi_3}=\frac{1}{\sqrt{2}}\left(\ket{\text{LG}_{+3,0}}+\ket{\text{LG}_{-3,0}}\right)$. (a) shows the real and imaginary parts of the density matrix for the undistorted state. In (b) we show the density matrix for the aberrated qubit. In this case, the strength of the simulated turbulence is characterized by $C_n^2$ = 80$\times 10^{-13} \text{mm}^{-2/3}$. In (c) we show the density matrix for the corrected qubit after applying our turbulence correction protocol.}
 \label{fig:4}
 \end{figure}

For sake of completeness,  we perform quantum state tomography of spatial modes. This allows us to certify the recovery of spatial coherence at the single-photon level. For this purpose we use superpositions of the following form, $\ket{\psi_{\ell}}=\alpha \ket{\text{LG}_{+\ell,0}}+\beta \ket{\text{LG}_{-\ell,0}}$, where $\alpha$, and $\beta$ represent complex amplitudes \cite{kwiat:2005,daniel:2001,adrien:2015,brandon:2016}. For simplicity, in our experiment we use the following spatial qubit $\ket{\psi_3}=\frac{1}{\sqrt{2}}\left(\ket{\text{LG}_{+3,0}}+\ket{\text{LG}_{-3,0}}\right)$. In Fig. \ref{fig:4}(a), \ref{fig:4}(b), and \ref{fig:4}(c), we show the real and imaginary parts of the reconstructed density matrices in the absence of turbulence, with turbulence, and after applying turbulence correction, respectively.
As shown in Fig. \ref{fig:4}(a), in this case, all the elements of the real part of the density matrix should be equal to 1/2, and the matrix elements of the imaginary part should be 0. The presence of any deviation from that is attributed to experimental imperfections. Furthermore, Fig. \ref{fig:4}(b) shows the detrimental effects produced by turbulence. The strength of turbulence in this case is $C_n^2$ = 80$\times 10^{-13} \text{mm}^{-2/3}$. Remarkably, after applying our machine learning protocol, we recover the original state almost perfectly. The shown density matrices certify the robustness of our technique.  We quantify the fidelity using $\mathscr{F}=\left(\text{Tr}\sqrt{\sqrt{\rho_{1}}\rho_{3} \sqrt{\rho_1}}\right)^2$, where $\rho_1$ and $\rho_3$ represent the density matrices of original and turbulence corrected spatial qubit. In our case, the experimental fidelity is 99.8\%.

\section{Conclusion}
\label{sec:4}
\ \ \ Spatial photonic modes have been in the spotlight for the past few decades due to their enormous potential as quantum information resources. However, these modes are fragile and vulnerable to random phase fluctuations induced by turbulence.  Unfortunately, these problems are exacerbated at the single-photon level. Indeed, the fragility of spatial modes of photons imposes important limitations on the realistic implementation of optical technologies in free-space. In this work, we have experimentally demonstrated the first smart communication protocol that exploits the self-learning features of convolutional neural networks to correct the spatial profile of single photons. This work represents a significant improvement over conventional schemes for turbulence correction \cite{roden:2014,hugo:2018}. The high fidelities achieved in the reconstruction of the spatial profile of single photons make our technique a robust tool for free-space quantum technologies. We believe that our work has important implications for the realistic implementation of photonic quantum technologies.

\section*{Acknowledgments}
\ \ \ N.B. would like to acknowledge the financial support from Army Research Office (ARO) under grant no. W911NF-20-1-0194. C.Y., J.F., and O.S.M.L. are supported by the U.S. Department of Energy, Office of Basic Energy Sciences, Division of Materials Sciences and Engineering under Award DE-SC0021069. R.T.G. acknowledges funding from the U.S. Office of Naval Research under grant number N000141912374. 

\bibliographystyle{vancouver.bst}
\bibliography{main.bib}
\end{document}